Rapid Communication

# Biferroic YCrO$_3$

Claudy Rayan Serrao[1,2], Asish K. Kundu[1], S. B. Krupanidhi[2], Umesh V. Waghmare[1,3], and C. N. R. Rao[1,2] *

[1]*Chemistry and Physics of Materials Unit, Jawaharlal Nehru Centre for Advanced Scientific Research, Bangalore –560064, India.*

[2]*Material Research Centre, Indian Institute of Science, Bangalore-560012, India.*

[3]*Theoretical Sciences Unit, Jawaharlal Nehru Centre for Advanced Scientific Research, Bangalore –560064, India.*

Abstract:

YCrO$_3$ which has a monoclinic structure, shows weak ferromagnetism below 140 K (T$_N$) and a ferroelectric transition at 473 K accompanied by hysteresis. We have determined the structure and energetics of YCrO$_3$ with ferromagnetic and antiferromagnetic ordering by means of first-principles density functional theory calculations, based on pseudopotentials and a plane wave basis. The non-centrosymmetric monoclinic structure is found to be lower in energy than the orthorhombic structure, supporting the biferroic nature of YCrO$_3$.

*Corresponding author: cnrrao@jncasr.ac.in

There has been much interest in biferroic materials in recent years.[1-4] Research in this field is of great interest because of the potential uses of materials with simultaneous ferroelectric and magnetic orderings. Some of these materials exhibit a ferroelectric transition at a relatively high temperature and a magnetic transition at a lower temperature. One such material in this category is $BiMnO_3$ with the ferroelectric transition ($T_E$) at 450 K and the ferromagnetic transition around ($T_C$) 105 K.[5,6] $BiFeO_3$, on the other hand, exhibits a $T_E$ of 1110 K and an antiferromagnetic transition ($T_N$) at 670 K.[7] $BiCrO_3$ is recently reported to exhibit a $T_E$ of 440 K and $T_N$ of 114 K.[8] Unlike these materials, $TbMnO_3$ shows a different behavior, wherein $T_E$ is lower than the $T_C$; here, the spin frustration causes the ferroelectric distortion. In addition to $BiMnO_3$, hexagonal rare earth manganates $LnMnO_3$ (with Ln = Ho, Er, Tm, Yb, Lu or Y) are known to be biferroic with $T_E \gg T_N$.[9] Thus, $YMnO_3$ is ferroelectric around 914 K and antiferromagnetic at 80 K.[10] Considering that $BiCrO_3$, is biferroic just like $BiMnO_3$, it occurred to us that rare earth chromites such as $YCrO_3$ may exhibit interesting biferroic properties, a view somewhat strengthened by the suspected ferroelectricity in some of them.[11]

Careful powder diffraction studies show that $YCrO_3$ has a monoclinic structure, as reported in the literature,[12] but the Rietveld analysis could not clearly distinguish between the centrosymmetric and non-centrosymmetric structures. However, our first-principles study (discussed later in the article) has shown the non-centrosymmetric structure to be energetically more stable. $YCrO_3$ is an antiferromagnet with weak ferromagnetism ($T_N$ = 140 K).[13,14] Our magnetic measurements show an increase in the magnetization around 140 K and the presence of magnetic hysteresis below this temperature (Fig. 1). Clearly, $YCrO_3$ exhibits features similar to those of a canted antiferromagnetic system.

We have investigated the dielectric properties of pressed pellets as well as thin films of $YCrO_3$. The thin films were deposited using a KrF excimer laser of 248nm (lambda physik) at 650°C in an oxygen ambient of 100mTorr with pulse energy of 140mJ/pulse on <111> oriented $Pt/TiO_2/SiO_2/Si$ substrates. The dielectric phase transition observed in a $YCrO_3$ pellet is shown in Figure 2(a). The transition occurs over the 410-440 K range with a maximum ($T_{max}$) of 418K at 500Hz. The $T_{max}$ is frequency dependent. The Curie Weiss plot of $1/\varepsilon$ vs T gives a $T_C$ of 473 K which is frequency independent as expected of a ferroelectric material.[15] The dielectric constant shows a large dispersion below $T_C$, but is frequency-independent above $T_C$, a behavior commonly observed in relaxor ferroelectrics.[16] The room-temperature dielectric constant is around 8000 at 5 kHz. A broad transition in normal ferroelectrics is attributed due to the presence of fine grains and the relaxor-like behavior may not be the property of the parent material itself. The presence of charged oxygen vacancies and the short range polar regions can, however, give rise to a relaxor like behavior in a normal ferroelectric.[17] The role of oxygen vacancies and the presence of nanopolar regions and their role in the relaxor-like behavior in $YCrO_3$ is under study.

Studies of the $YCrO_3$ thin films, show a dielectric phase transition around 400K at 500 Hz as can be seen from Fig. 2(b). The monotonous increase in the dielectric constant at low frequencies above the $T_C$ is associated with the dc conduction. Accordingly, we observe high dissipation factors of 2.348 and 38.834 (5 kHz) at room temperature in the bulk sample and the thin film respectively. The high-temperature loss shows a monotonous rise due to the increase in the dc leakage current at high temperatures, and efforts are in progress to minimize the losses through processing. The dielectric phase transition found both in the pellets and the thin films also lends evidence for the presence of a ferroelectric phase in $YCrO_3$.

The ferroelectric behavior of the $YCrO_3$ pellets and films was also confirmed by their room temperature capacitance-voltage (C-V) characteristics. The butterfly nature of the CV curves [Fig. 3(a)] suggests a weak ferroelectric behavior at room temperature.[18] The room-temperature polarization-electric field (P-E) measurements on the pellets and thin films of $YCrO_3$ show the presence of hysteresis. We show hysteresis curves at different temperatures in Fig. 3(b). The maximum polarization observed is 2 $\mu C/cm^2$ at 300 K in the case of the pellet and 3 $\mu C/cm^2$ at 178K in the case of the thin film. Clearly, $YCrO_3$ exhibits a weak ferroelectric behavior from $T_E$ to low temperatures through the magnetic transition.

Based on measurements of the X-ray powder diffraction data, $P2_1/n$ (space group no.14) has been suggested as a probable space group for $YCrO_3$ in the literature.[12] As $P2_1/n$ is a centrosymmetric monoclinic space group, our finding of ferroelectricity in $YCrO_3$ needs an explanation. To this end, we have carried out first-principles spin dependent density functional theory calculations using a standard plane-wave code PWSCF 2.0.1[19] with a generalized gradient approximation[20] to the interaction energy of electrons. We used ultra-soft pseudo-potentials[21] to represent the interaction between the ions and the electrons and treated semi-core $s$ and $p$ states of Y and Cr explicitly. An energy cutoff of 30 Ry (180 Ry) on the plane wave basis was used in representation of wave-functions (density). Most calculations involved unit cells with 40 and 20 atoms, and Brillouin zones integrations were sampled with a Monkhorst-Pack mesh of *k*-points that is equivalent to a 4×4×4 uniform mesh[22] in the Brillouin zone of a crystal with 5 atoms per cell. In all calculations, we used experimental lattice parameters or volume of unit cell.[23]

As $P2_1/n$ is a relatively low-symmetry space group, determination of the YCrO$_3$ structure with many structural parameters becomes challenging. We followed two procedures: (a) We first optimized the structure with a constraint of $P2_1/n$ symmetry. A further relaxation was carried out by breaking its inversion symmetry through off-centering of Y and Cr atoms. (b) We started with an initial structure obtained by randomly off-centering various atoms in the perovskite unit cell, and relaxed the structure to minimize energy. Structural optimization was first carried out maintaining a *G*-type antiferromagnetic ordering, and it was reassuring to find essentially the same final structure (Fig. 4) with both the procedures. The final structure has a broken inversion symmetry, supporting the possibility of ferroelectricity in YCrO$_3$. From the procedure (a), we find that the final AFM structure is indeed lower in energy than the relaxed centrosymmetric $P2_1/n$ phase by 0.025 eV/formula unit. Our guess for the space group of the lowest energy structure of YCrO$_3$ is the monoclinic $P2_1$ (no. 4), though we find tiny distortions of the structure that amount to further lowering of symmetry. In this distorted perovskite structure, the average oxygen coordination of Y is about 6. Structural distortions of the parent cubic perovskite structure leading to the ferroelectric structure [shown in Figure 4(a)] have been analyzed by writing them as a linear combination of normal modes of the cubic structure. For YCrO$_3$, we find the largest component of structural distortions to be associated with (a) $M_3$ modes [k = (011) π/a], (b) $R_{25}$ modes [k

= (111) π/a], which correspond to rotations of oxygen octahedral, and (c) $M_5$ modes. There is a small component of modes with Y and oxygen displacements such as $X_5'$, $R_{15}$ and $\Gamma_{15}$ modes. There are no structural distortions with Cr displacements. The modes in (a) and (b) dominating the low-temperature phase do not break the inversion symmetry, while the relatively weak distortions of the mode $\Gamma_{15}$ give rise to ferroelectricity.

Using the relaxed structure with AFM ordering as an initial guess, we optimized the structure of $YCrO_3$ with ferromagnetic ordering. While the relaxed structure with ferromagnetic ordering also exhibits a broken inversion symmetry, it is higher in energy than the structure with *G*-AFM ordering by about 0.04 eV/formula unit. For both types of ordering, the self-consistently determined local magnetic moment at Cr site is close to 3 $\mu_B$ consistent with fully occupied $t_{2g}$ states of Cr just below the energy gap, as seen in the densities of electronic states (shown in Fig. 5). This results in a spin density with almost cubic symmetry centered at the Cr sites [see Fig. 4(b)]. Unlike many other materials, $YCrO_3$ is insulating in both FM and AFM ordered states with an energy gap of 1.3 eV and 1.8 eV respectively, and exhibits a large exchange splitting. We used the Berry phase[24] method to compute polarization and find *P* along a-axis to be about 3 $\mu C/cm^2$.

To understand the origin of ferroelectricity in $YCrO_3$ we studied the structural instabilities in the cubic perovskite structure of $YCrO_3$. We used a unit cell doubled along the <111> direction with a *G*-type AFM ordering and determined phonon frequencies at its Γ point, which correspond to phonons at Γ and *R* (folded for the supercell) points of the single perovskite unit cell. We find two triply degenerate instabilities, the strongest one (about 300 $cm^{-1}$) exhibits rotational motion of oxygen octahedra, and a weaker instability (about 100 $cm^{-1}$) of Y displacements with respect to oxygen cage. Note that the former gives an anti-ferrodistortive phase while the latter is responsible for a ferroelectric phase. To probe anharmonic strength of the ferroelectric instability, we distorted the structure by freezing in Y displacements and relaxed it maintaining a rhombohedral symmetry and *G*-AFM ordering. We find an energy lowering by 0.089 eV/formula unit, comparable to other perovskite ferroelectrics like $PbTiO_3$.[25] Interestingly, for a rhombohedral symmetry and FM ordering, optimized ferroelectric structure is lower than the paraelectric structure by 0.13 eV/formula unit, though is higher in energy than the ferroelectric structure with AFM ordering. In both FM and AFM ordering, off-centering of Cr atoms is not favored energetically.

In conclusion, our experimental results clearly demonstrate $YCrO_3$ to be biferroic, with interesting magnetic properties. Our calculations support the observed ferroelectricity in $YCrO_3$ through the determination of the detailed structure. There are competing structural instabilities in $YCrO_3$, and the dominating one is of anti-ferrodistortive type. Hence, the polarization found in $YCrO_3$ arising from the weak ferroelectric instability is relatively small. Our finding that the *G*-type AFM ordering is lower in energy than the uniform FM ordering is consistent with the experimentally known AFM phase of $YCrO_3$ and indicates that ferromagnetism in $YCrO_3$ can only be weak and its emergence at lower temperatures must be from mechanisms other than simple exchange type magnetic interactions, possibly canted AFM. It is possible that non-centrosymmetry in $YCrO_3$ is local in nature, an aspect that can be revealed by a study of the pair-distribution functions based on careful diffraction measurements. It appears likely that other chromites of heavier rare earths (Er, Ho, Yb, Lu) may also be biferroic.

Our preliminary measurements show that LuCrO$_3$ becomes ferromagnetic at 115 K (T$_N$) and ferroelectric at 488 K (T$_E$). Further studies are in progress, on these chromites.

The authors thank the Department of Science and Technology for support of this research. UVW thanks J. Bhattacharjee for help with xcrysden.

**Figure captions**

Fig. 1  Temperature-variations of magnetization of YCrO$_3$ (pellet). Inset shows magnetic hysteresis curves at different temperatures.

Fig. 2  Temperature-variations of dielectric constant of YCrO$_3$ (a) pellet and (b) a thin film at different frequencies.

Fig. 3  (a) Capacitance-voltage curves of YCrO$_3$ (b) dielectric hysteresis in YCrO$_3$ at different temperatures.

Fig. 4  Structure of AFM YCrO$_3$ (8 formula units/cell): (a) structural dirtortions of the parent centrosymmetric structure are shown with arrows with lengths proportional to atomic displacements from the cubic structure, (b) cube-like surfaces surrounding Cr atoms are the isosurfaces of magnetization density at a value 10 % of its maximum. We have used XCRYSDEN software for visualization.[26]

Fig. 5  Density of electronic states with spin up (positive) and down (negative) near the energy gap of monoclinic structures with (a) antiferromagnetic and (b) ferromagnetic ordering.

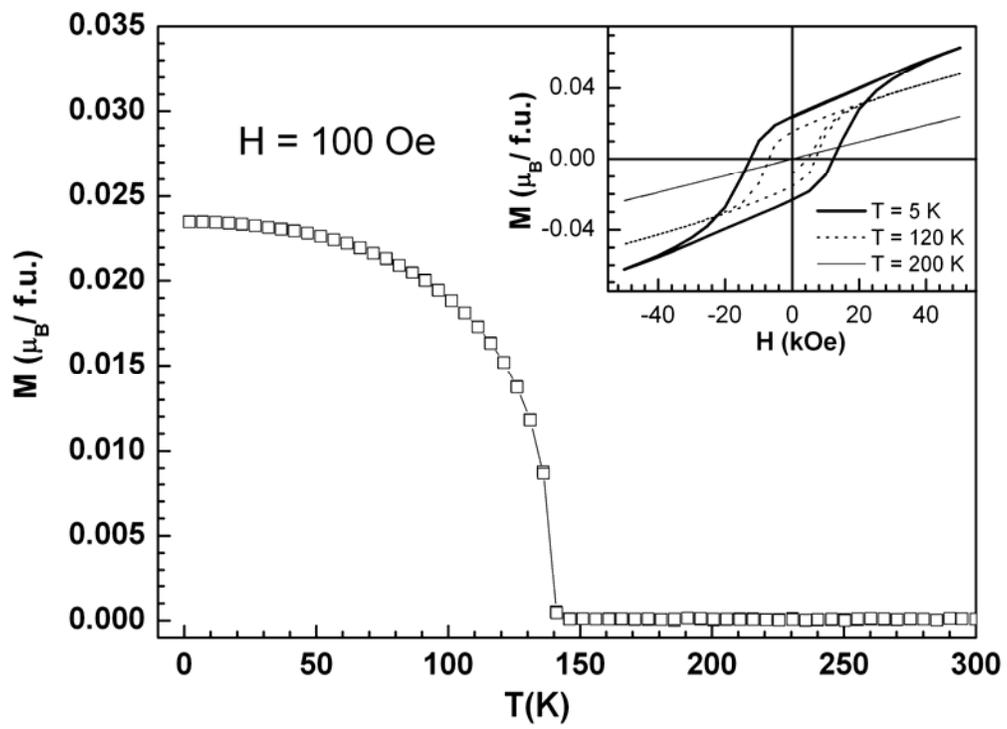

**Fig. 1**

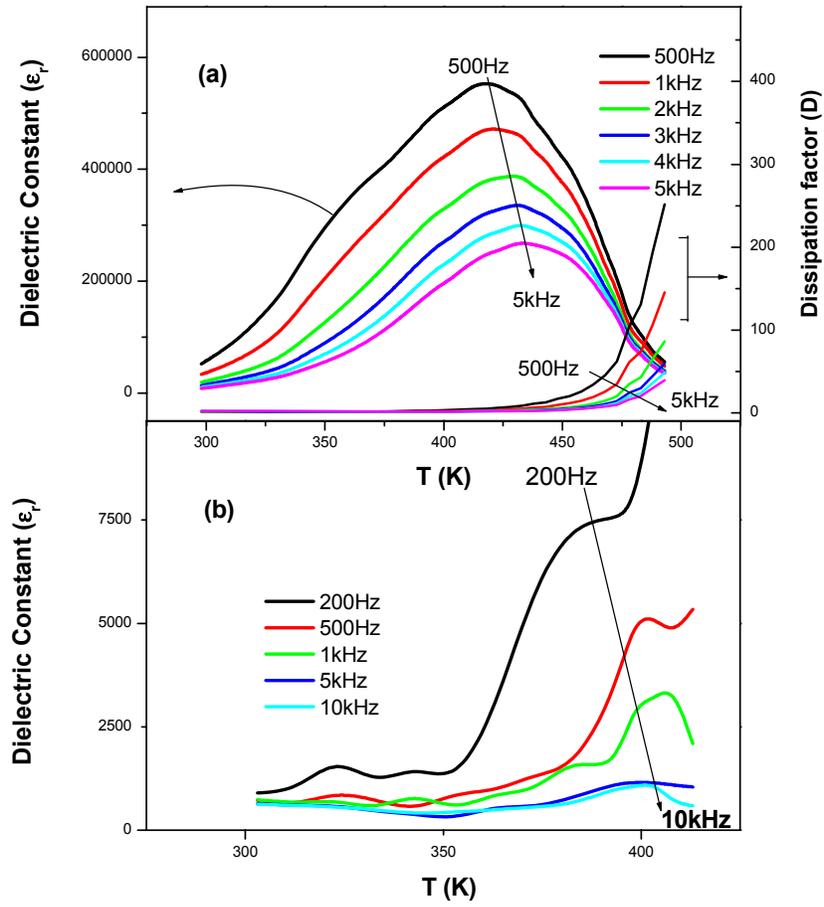

**Fig. 2**

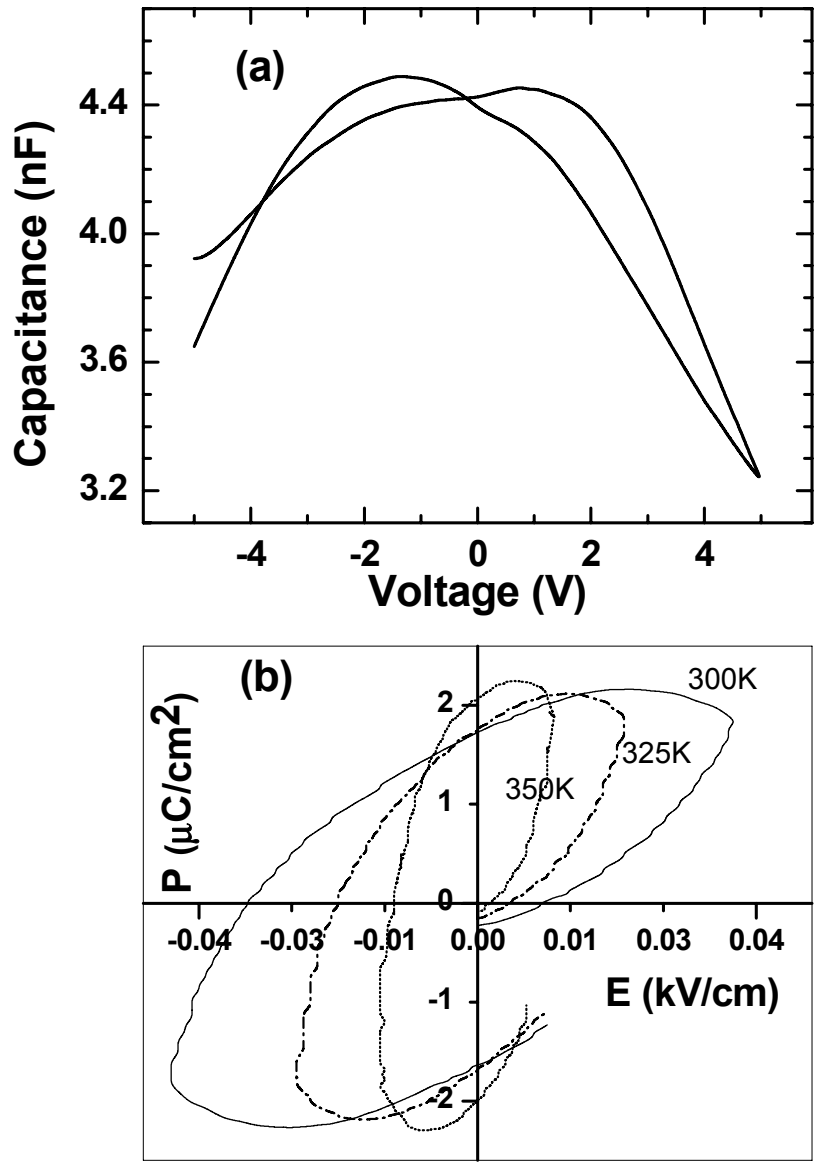

**Fig. 3**

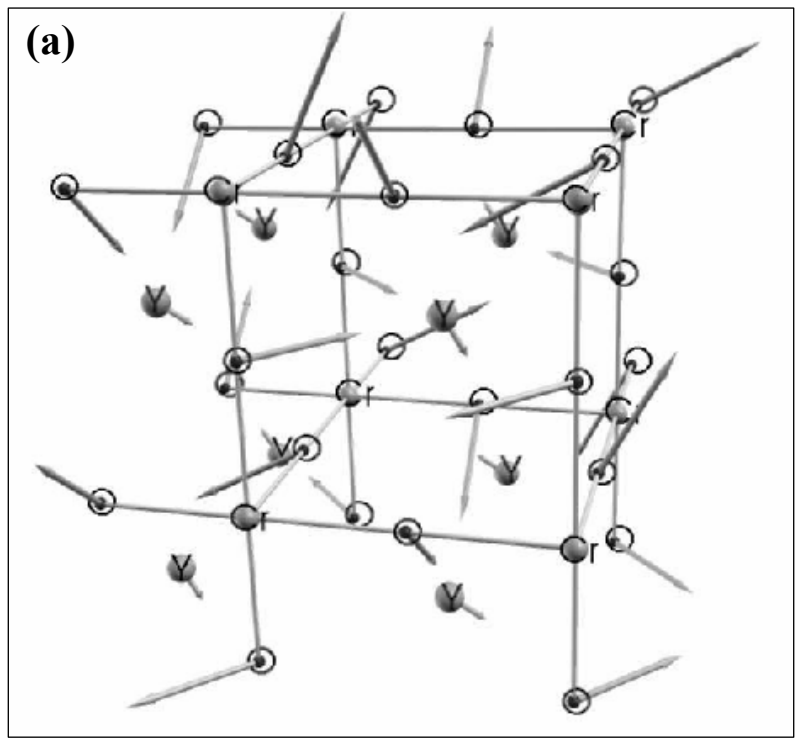

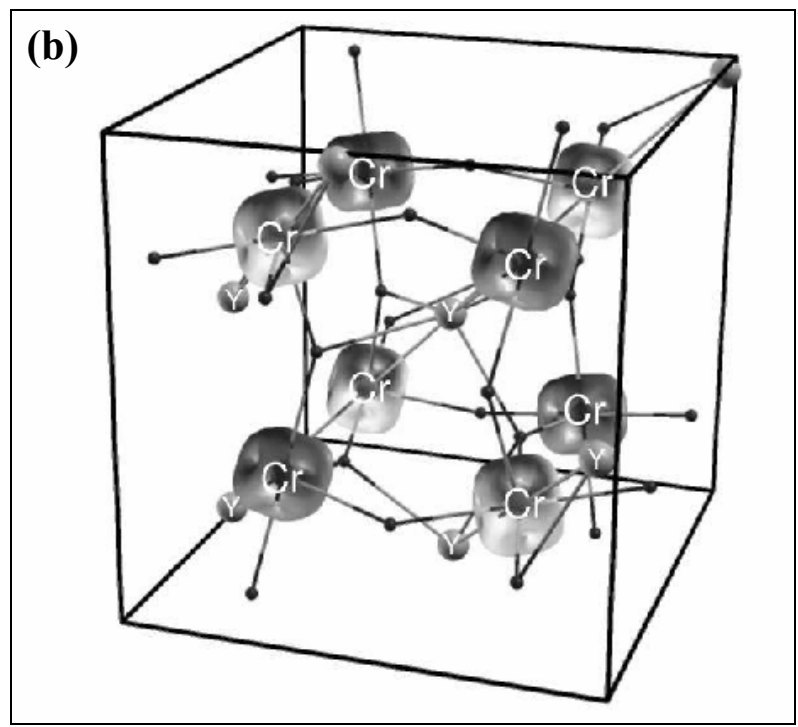

**Fig. 4**

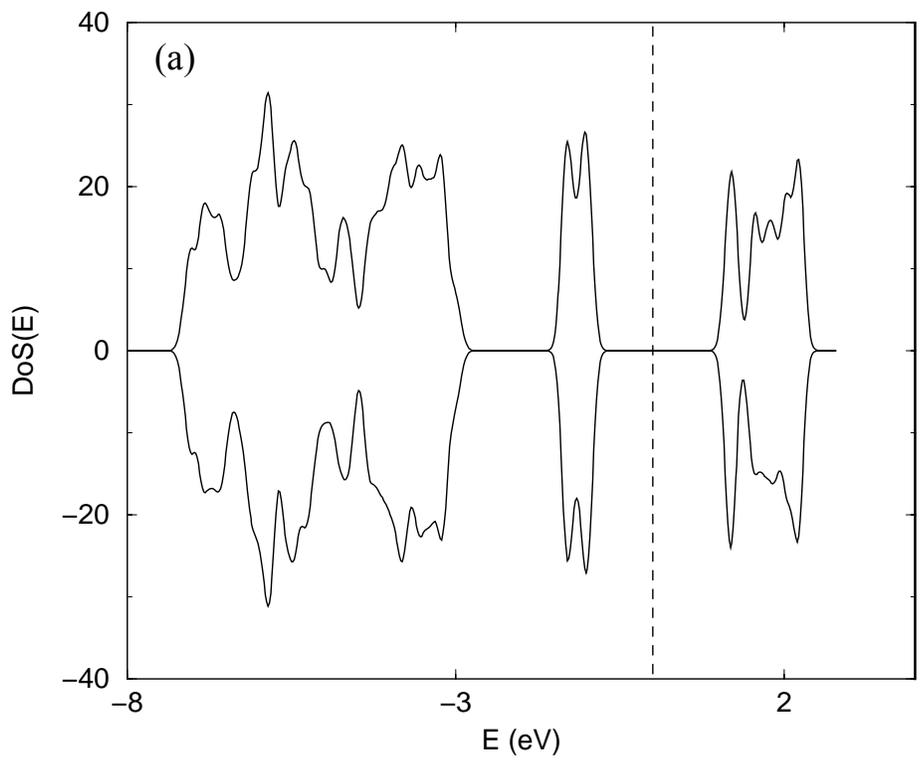

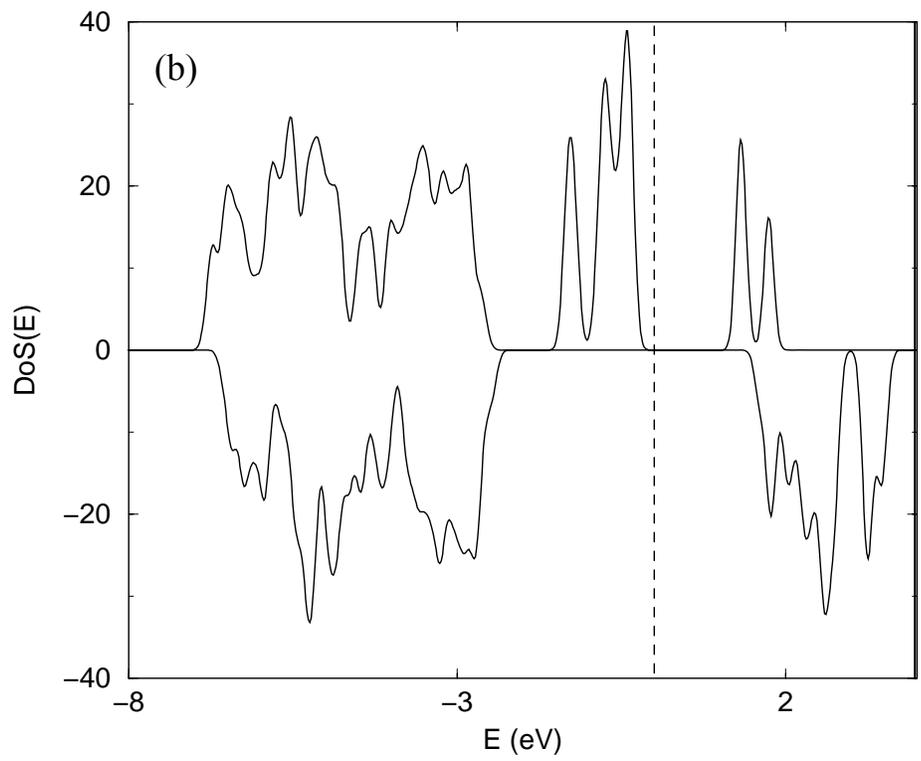

**Fig. 5**